# A Memory Hierarchical Layer Assigning and Prefetching Technique to Overcome the Memory Performance/Energy Bottleneck[*]

Minas Dasygenis
mdasyg@ee.duth.gr

Erik Brockmeyer
Erik.Brockmeyer@imec.be

Bart Durinck
Bart.Durinck@imec.be

Francky Catthoor
Francky.Catthoor@imec.be

Dimitrios Soudris
dsoudris@ee.duth.gr

Antonios Thanailakis
thanail@ee.duth.gr

## Abstract

*The memory subsystem has always been a bottleneck in performance as well as significant power contributor in memory intensive applications. Many researchers have presented multi-layered memory hierarchies as a means to design energy and performance efficient systems. However, most of the previous work do not explore trade-offs systematically. We fill this gap by proposing a formalized technique that takes into consideration data reuse, limited lifetime of the arrays of an application and application specific prefetching opportunities, and performs a thorough trade-off exploration for different memory layer sizes. This technique has been implemented on a prototype tool, which was tested successfully using nine real-life applications of industrial relevance. Following this approach we have able to reduce execution time up to 60%, and energy consumption up to 70%.*

## 1. Description of our work

We overcome the performance and energy bottleneck that memory imposes at the system by developing the Memory Allocation and Layer Assignment [1] with time extensions (MHLA with TE) technique. Our technique targets at the data reuse opportunities, found in most multimedia and image processing applications, array in-place optimizations and optimum memory transfer scheduling. By exploiting data reuse, a part of an array is copied from one layer to a lower layer, closer to the processor. As a result, energy and performance can be improved since most accesses take place on the smaller copy and not on the large more energy consuming and slower higher memory layer. TE enables us to increase the performance of a system, with minimal cost. They enable the selective prefetching of copy candidates from off-chip memory layers to on-chip memory layers, exploiting the lifetime information of copies. Time extensions, need the support of a memory transfer engine (like DMA engine or data mover) that allows simultaneous the CPU to continue processing data and the engine to copy off-chip data to on-chip layers. The goal of time extensions is to hide as much as possible the cycles required in accessing off-chip memory, respecting data dependencies and on-chip size requirements. In case that our architecture does not support a memory transfer engine, TE are not applicable.

## 2. MHLA and Time Extensions

The MHLA technique with time extensions has been implemented in a tool with the same name (MHLA), which is able to find all the optimal trade-off points, given some architecture specific constraints and models. The exploration flow can be divided into two distinct steps: A selection and assignment step and a time extension step. Every step has a number of substeps. When we first presented the MHLA technique [1], it did not consider TE. We present a new version of our MHLA technique with the major contribution the support of TE. In our technique TE is the second step in MHLA.

Time extensions are done in an iterative process (Figure 1). We examine every DMA Block Transfer (BT) and we try to schedule earlier the initiating of the DMA, obeying dependencies and on-chip memory size.

We iterate over the list of BTs in the greedy order and try to perform prefetching. We evaluate whether the increase in lifetime of the copy candidate, due to the extension of the DMA one loop before, is valid or not. If the increase in lifetime increases on-chip memory size over the user-defined on-chip memory constraint, then this extension is not valid and no further actions are performed for this BT. Otherwise, TE for one loop before is valid for this BT.

```
foreach BT(i) {
 if ( is_DMA(BT(i)) ) {
   BT_list+=BT(i);
   /* Estimate Cycles */
   BT_time(i)=compute_time(BT(i));
   BT_sort_factor(i)=
           BT_time(i)/size(BT(i));
   deps=dep_analysis(BT(i));
   BT_freedom_loops(i)=
        loops_between(deps, BT(i));
}}
sort(BT_list, BT_sort_factor);
foreach BT(i) in BT_list {
/* Initially no TE */
extended_cycles=0;
forearch loop in BT_freedom_loops(i) {
   if (fits_size(BT(i), loop)) {
     /* Take next BT */
     break; }
   cpu_cycles=compute_loop_cycles();
   ext_cycles+=cpu_cycles;
   if (ext_cycles>=BT_time(i)) {
     /* Fully time extended */
     break; }
 }
 dma_priority();
```

**Figure 1. The TE step performs an application specific prefetching.**

## 3. Experimental Results

The usefulness and feasibility of the MHLA technique is demonstrated using nine real life applications of motion estimation, video encoding, image and audio processing domain. Although, we only consider single threaded applications, we plan to extend our technique to multiple tasks with multiple threads.

MHLA technique performs exploration in two steps. After deciding and placing on memory layers, arrays and copies the step of time extensions is applied. On the applications considered, the first step boost performance from 40% to 60% compared to the out of the box code (Figure 2) for specific memory sizes. An optimum memory allocation and assignment can also reduce energy consumption significantly up to 70% (Figure 3).

The MHLA second step tries to perform prefetching on the copies that have been decided in the previous step. This step can boost performance of up 33%, if there are a lot of processing loops that can hide prefetching block transfers. This step pushes performance towards the ideal case where ever block transfer can be hidden from the processor (0 wait cycles block transfer time) (Figure 2). Energy consumption in both steps remains the same, because in our models we only consider accesses to the memory hierarchy. Though, it is expected that the reduced execution time will result in lower energy consumption.

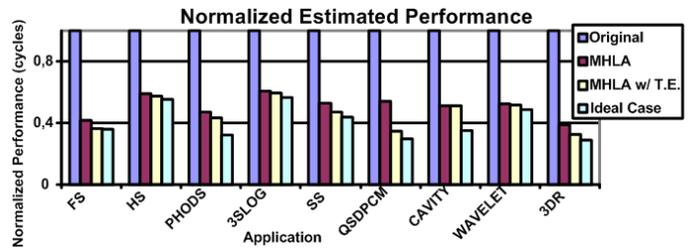

**Figure 2. MHLA improves performance up to 60%. MHLA with TE can boost performance even more.**

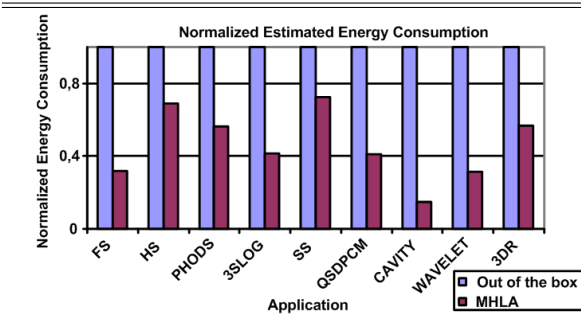

**Figure 3. In addition to performance improvements, MHLA technique benefits energy consumption as well.**

## 4. Conclusions

We have addressed the performance/energy consumption bottleneck problem by developing a a formalized technique of memory allocation and assignment. This technique, exploits data-reuse opportunities, in-place optimizations and application specific prefetching, which should be used during the early system design steps. This technique has been implemented in a prototype tool, which is part of the ATOMIUM framework, and has allowed us to do fast, accurate and automatic exploration of nine real-life applications. We found significant performance and energy consumption gains on every application, illustrating the importance of our technique in the early design phases.